\documentclass[rmp,nofootinbib,amsmath,amssymb,twocolumn]{revtex4}
\usepackage[T1]{fontenc} \usepackage[latin1]{inputenc}
\usepackage{amsmath} \usepackage{graphicx} \usepackage{amssymb}

\makeatletter

\usepackage{subfigure} \usepackage{amsfonts}

\makeatother
\begin{document}

\title{Passage Times for Polymer Translocation Pulled through a Narrow
Pore}

\author{Debabrata Panja} \affiliation{Institute for Theoretical
Physics, Universiteit van Amsterdam, Valckenierstraat 65, 1018 XE
Amsterdam, \\The Netherlands} \author{Gerard
T. Barkema$^{\dagger\ddagger}$} \affiliation{$^{\dagger}$Institute for
Theoretical Physics, Universiteit Utrecht, Leuvenlaan 4, 3584 CE
Utrecht, The Netherlands\\ $^{\ddagger}$Instituut-Lorentz,
Universiteit Leiden, Niels Bohrweg 2, 2333 CA Leiden, The Netherlands}

\begin{abstract}
We study the passage times of a translocating polymer of length $N$ in
three dimensions, while it is pulled through a narrow pore with a
constant force $F$ applied to one end of the polymer. At small to
moderate forces, satisfying the condition $FN^{\nu}/k_BT\lesssim1$,
where $\nu\approx0.588$ is the Flory exponent for the polymer, we
find that $\tau_N$, the mean time the polymer takes to leave the pore,
scales as $N^{2+\nu}$ independent of $F$, in agreement with our
earlier result for $F=0$. At strong forces, i.e., for
$FN^{\nu}/k_BT\gg1$, the behaviour of the passage time crosses over to
$\tau_N\sim N^2/F$. We show here that these behaviours stem from the
polymer dynamics at the immediate vicinity of the pore --- in
particular, the memory effects in the polymer chain tension imbalance
across the pore.
\end{abstract}

\maketitle

\section{Introduction\label{sec1}}
Molecular transport through cell membranes is an essential mechanism
in living organisms. Often, the molecules are too long, and the pores
in the membranes too narrow, to allow the molecules to pass through as
a single unit. In such circumstances, the molecules have to deform
themselves in order to squeeze --- i.e., translocate --- themselves
through the pores. DNA, RNA and proteins are such naturally occuring
long molecules \cite{drei,henry,akimaru,goerlich,schatz} in a variety
of biological processes. Translocation is also used in gene therapy
\cite{szabo,hanss}, in delivery of drug molecules to their activation
sites \cite{tseng}, and as a potentially cheaper alternative for
single-molecule DNA or RNA sequencing
\cite{expts1,nakane}. Consequently, the study of translocation is an
active field of research: as a cornerstone of many biological
processes, and also due to its relevance for practical applications.

Translocation in living organisms is a complex process. Take for instance
the case of gene expression: most proteins are synthesized within the
cytoplasm. Their subsequent accurate and swift delivery to target
sites, requiring energy, is a crucial step in gene expression. In
different situations the energy is provided by chaperon molecules
\cite{buchner}, pH gradient \cite{magzoub} or molecular motors across
membranes \cite{wuite}. These delivery mechanisms can be further
complicated by membrane fluctuations and sometimes by gates that
control the accessibility of the pores \cite{gate}. In view of such
complexity, translocation as a biological or biophysical process in
living organisms has been scrutinized in a variety of {\it in vivo\/}
experimental situations [see e.g. Ref. \cite{nelson} and the
references therein].

The experimental developments have been followed by a number of 
mean-field type theoretical studies on polymer translocation
\cite{theory}.

More recently, translocation has found itself at the forefront of
single-molecule-detection experiments \cite{nakane,expts}, as new
developments in design and fabrication of nanometer-sized pores and
etching methods may lead to cheaper and faster technology for the
analysis and detection of single macromolecules.  The underlying
principle for these experiments is that of a Coulter counter:
molecules suspended in an electrolyte solution pass through a narrow
pore in a membrane. The electrical impedance of the pore increases
with the entrance of a molecule as it displaces its own volume of the
electrolyte solution. By applying a voltage over the pore, the passing
molecules are detected as current dips. For nanometer-sized pores
(slightly larger than the molecule's cross-section) the magnitude and
the duration of these dips have proved to be effective in determining
the size and length of the molecules. In the case of DNA sequencing at
nucleotide level, usage of protein pores (modified
$\alpha$-haemolysin, mitochondrial ion channel, nucleic acid
binding/channel protein etc.), and etching specific DNA sequences
inside the pores \cite{szabo,proteinpore} have opened up promising new
avenues of fast, simple and cheap technology for single macromolecule
detection, analysis and characterization [see Ref. \cite{mass} for a
recent development].

The subject of this paper is a translocating polymer threaded through
a narrow pore in an immobile membrane, where a bead is attached to one
end of the polymer, and the bead is pulled by an optical tweezer with
a constant force. Such a setup can be used to spread apart a partially
unzipped dsDNA molecule --- of which one strand is threaded through
the pore --- a process that can quantify the forces involved in
basepair unzipping kinetics \cite{mathe}. In theoretical literature,
this problem has been considered in recent times: for polymer length
$N$ and applied force $F$, in Ref. \cite{kantor}, in the {\it
absence\/} of hydrodynamical interactions, a lower bound $\propto N^2$
for small forces ($FN^\nu/k_BT\le1$) has been argued for the polymer's
mean unthreading time $\tau_N$, the average time it takes  for the
polymer to leave the pore. The lower bound holds in the limit of
unimpeded polymer movement, i.e., for an infinite pore, or
equivalently, in the absence of the membrane. Simulation data (in
two-dimensions) presented in Ref. \cite{kantor} indicated that the
lower bound may very well be valid in the limit of narrow pores as
well. The same problem, also in the {\it absence\/} of hydrodynamical
interactions, has been numerically studied in two-dimensions in
Ref. \cite{luo}. It reported that for narrow pores $\tau_N\sim N^2$
with the velocity of translocation $v(t)\sim N^{-1}$ for moderate and
strong forces. As the force-dependence of $\tau_N$ is concerned,
Ref. \cite{luo} reported numerical results that in the absence of the
membrane $\tau_N\sim F^{-2+\frac{1}{\nu}}$ for moderate forces, and
$\tau_N\sim F^{-1}$ for strong forces, while for narrow pores
$\tau_N\sim F^{-1}$ for moderate to strong forces.

The purpose of this paper is to revisit the problem of translocation
of a polymer pulled through a narrow pore in the {\it absence\/} of
hydrodynamical interactions, in order to provide a deeper theoretical
understanding of the polymer dynamics under these conditions, as well
as of the scaling behaviour of the unthreading time of the polymer. In
support of our theory, we perform high precision computer simulations,
using a {\it three-dimensional self-avoiding lattice polymer model\/}
that we have used before to  study polymer translocation
\cite{wolt,anom,anomlong} and several other situations
\cite{others}. Our conventions to study this problem, all throughout
this paper, is the following. We place the membrane at $z=0$, and
thread the polymer of {\it total length\/} $2N$ halfway through the
pore such that both the right ($z>0$) and the left ($z<0$) of the
membrane have equal number of monomers $N$. We fix the middle monomer
(monomer number $N$) at the pore, apply a force $F$ on the free end on
the right and let both left and right segments of the polymer come to
equilibrium. At $t=0$ we release the middle monomer
and let translocation commence. The mean time $\tau_N$ that the
polymer remains within the pore is defined as the mean unthreading
time for polymer length $N$ under the force $F$. Additionally, we use
$k_BT=1$, although $k_BT$ is explicitly mentioned at several places in
the paper.

Our main results in this paper are as follows. At small to moderate
forces, satisfying the condition $FN^{\nu}\lesssim1$, where
$\nu\approx0.588$ is the Flory exponent for the polymer, we find that
$\tau_N$ is independent of $F$. In agreement with our earlier result
for unbiased polymer translocation; i.e., for $F=0$ \cite{anom},
$\tau_N$ scales with polymer length as $\tau_N\sim N^{2+\nu}$.  At
strong forces, i.e., for $FN^{\nu}\gg1$, we find $\tau_N\sim
N^2/F$. While these results agree with the existing ones
\cite{kantor,luo} in broad terms, we show that $v(t)$, the velocity of
translocation is {\it not\/} constant in time. In fact, for strong
forces, we show  that the velocity of translocation $v(t)$ behaves as
$t^{-1/2}$, while for small to moderate forces the behaviour of $v(t)$
is more complicated. The physical picture provided in
Refs. \cite{kantor,luo}, wherein  the scaling arguments for the
unthreading time involved a constant velocity of translocation (albeit
an average one, in light of this work) is incomplete \cite{evi}. Using
theoretical analysis supported by high-precision simulation data, we
show that these behaviours stem from the dynamics of the polymer
segments at the immediate vicinity of the pore --- in particular, the
memory effects in the polymer chain tension imbalance across the
pore. The theoretical analysis presented here is based on that of
Ref. \cite{anom}, and therefore provides a direct confirmation of the
robustness of the theoretical method presented in Ref. \cite{anom}.

This paper is organized in the following manner. In Sec. \ref{sec2} we
discuss a method to measure component of the polymer chain tension
which is perpendicular to the membrane. In Sec. \ref{sec3a} we analyze
the memory effects in $\phi(t)$, the imbalance of this component of
the polymer chain tension.  In Sec. \ref{sec3} we discuss the
consequence of these memory effects on the translocation velocity
$v(t)$, and obtain the relation between the mean unthreading time
$\tau_N$ and the polymer length $N$. We finally end this paper with a
discussion in Sec. \ref{sec4}.

\section{Chain tension perpendicular to the membrane\label{sec2}}

A translocating polymer should be thought of as two segments of
polymers threaded at the pore, while the segments are able to exchange
monomers between them through the pore. In Ref. \cite{anom} we
developed a theoretical method to relate the dynamics of translocation
to the imbalance of chain tension between these two segments across
the pore. The key idea behind this method is that the exchange of
monomers across the pore responds to $\phi(t)$, this imbalance of
chain tension; in its turn, $\phi(t)$ adjusts to $v(t)$, the transport
velocity of monomers across the pore. Here, $v(t)=\dot{s}(t)$ is the
rate of exchange of monomers from one side to the other, where $s(t)$
is the total number of monomers transferred from one side of the pore
to the other in time $[0,t]$. In fact, we noted that $s(t)$ and
$\phi(t)$ are conjugate variables in the thermodynamic sense, with
$\phi(t)$ playing the role of the chemical potential difference across
the pore.

By definition, $\phi(t)=\Phi_R(t)-\Phi_L(t)$ where $\Phi_R(t)$ and
$\Phi_L(t)$ are respectively the chain tension (or the chemical
potential) on the right and the left side of the pore. Consider a
separate problem, where we tether one end of a polymer to a fixed
membrane, yet the number of monomers are allowed to spontaneously
enter or leave the tethered end, then we have
\begin{eqnarray}
\frac{W_t(-\rightarrow +)}{W_t(+\rightarrow-)}=\exp[\Phi(t)/k_BT]\,,
\label{e1}
\end{eqnarray}
where $W_t(-\rightarrow +)$ [resp. $W_t(+\rightarrow-)$] is the rate
that a monomer enters (resp. leaves) the polymer chain through the
tethered end at time $t$. Note that tethering the polymer while
allowing monomers to enter or leave the polymer at the tethered end is
precisely the case that translocation represents.

Returning to our problem of a translocating polymer under a pulling
force $F$, note that at $t=0$, when the left and the right segments
are equilibrated with $F=0$ and $F\neq0$ respectively, it
is easy to use Eq. (\ref{e1}) to measure the chain tension for both
segments at the pore [$\Phi(t=0)$ in our notation], since under these
conditions, we also have the relation that
\begin{eqnarray}
P_-\,W_{t=0}(-\rightarrow +)=P_+\,W_{t=0}(+\rightarrow-)\,,
\label{e2}
\end{eqnarray}
where $P_-$ (resp. $P_+$) is the probability that the (left or the
right) polymer segment has one monomer less (resp. one extra
monomer). Equations (\ref{e1}) and (\ref{e2}) together yield us
\begin{eqnarray}
\Phi(t=0)=k_BT\,\ln\frac{P_+}{P_-}\,.
\label{e3}
\end{eqnarray}

Note that even for $F=0$, there is nonzero chain tension, due to the
presence of the membrane. A polymer's free energy close to a membrane is
higher than its free energy in bulk. In other words, the membrane repels
the polymer, and as a result, for a polymer with one end tethered to
a membrane, the monomers close to the membrane are more stretched than
they would be in the bulk.

\begin{figure}[!h]
\hspace{-2.3cm}
\includegraphics[width=0.75\linewidth]{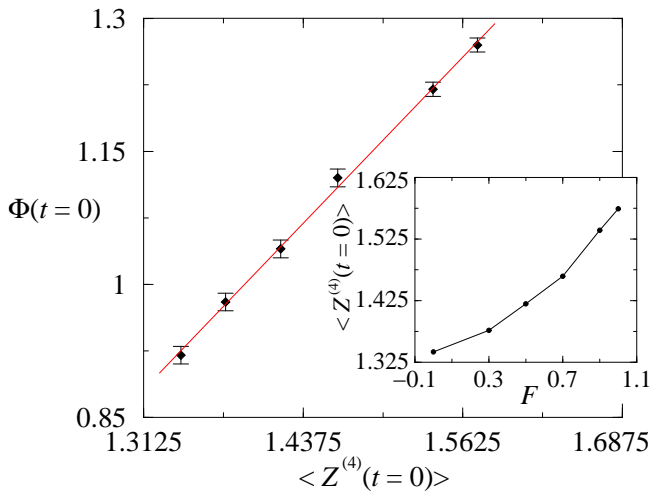}
\caption{$\langle Z^{(4)}(t=0)\rangle$ vs. $\Phi(t=0)$ demonstrating
the linear relationship between the two, for $N=100$ and
$F=0.0,0.3,0.5,0.7,0.9$ and $1.0$ respectively. The angular brackets
for $\langle Z^{(4)}(t=0)\rangle$ indicates an average over
$12,800,000$ polymer realizations. The data for
for $\Phi(t=0)$ are obtained $2,400$ polymer realizations. The red
line corresponds to the linear best-fit. Inset:
$\langle Z^{(4)}(t=0)\rangle$ as a function of $F$.\label{fig1}}
\end{figure}

The chain tension as obtained from Eq. (\ref{e3}) is linearly related
to the $z$-coordinate of the centre-of-mass of the first few monomers
along the polymer's backbone, at the immediate vicinity of the pore, at
least for the relatively modest forces used in our simulations. This is
shown in Fig. \ref{fig1}, where for a tethered polymer of length $N=100$,
the average distance $\langle Z^{(4)}(t=0)\rangle$ of the centre-of-mass
of the first $4$ monomers along the polymer's backbone, counting from
the tethered end of a polymer, is plotted versus the chain tension
$\Phi$, while its free end is pulled with various force strengths $F$.
Within the error bars, all the points in Fig. \ref{fig1} fall on a
straight line, implying that $\Phi$ is very well-proxied by $\langle
Z^{(4)}\rangle$. Note in Fig. \ref{fig1} that the black line does
not pass through the origin, which shows that $\Phi_{F=0}\neq0$, as we
argued above.  Since measurements of the chain tension via Eq. (\ref{e3})
are much more noisy than measurements of $\langle Z^{(4)}\rangle$,
we will use the latter quantity as a measure for the chain tension.

\section{Memory effects in the $z$-component of the chain tension
  \label{sec3a}} 

In the case of unbiased polymer translocation, we have witnessed in
Ref. \cite{anom} that the memory effects of the polymer gives rise to
anomalous dynamics of translocation. We argued \cite{anom} that the
imbalance of the chain tension $\phi(t)$ across the pore and the
number of monomers $s(t)$ that have crossed from one side of the
membrane to the other in time $[0,t]$ are conjugate variables in the
thermodynamic sense. Additionally, $\phi(t)$ is related to the
translocation velocity $v(t)$ by the relation
$\phi(t)=\phi_{t=0}-\int_{0}^{t}dt'\mu(t-t')v(t')$ via the memory
kernel $\mu(t)$, which can be thought of as the `impedance' of the
system. On average, there will not be an imbalance in chain tension if
no force is applied, but there will be fluctuations in chain tension.
When the polymer is pulled by a force $F$ to the right, the symmetry
between the polymer segments on two sides of the membrane (viz., the
polymer segments on the right are more stretched than those on the
left of the membrane), are destroyed. As a result, on average
$\phi(t)$, $\phi_{t=0}$ and $v(t)$ are non-zero, and from now on, we
understand these three quantities as an average over all the
unthreading polymers. Additionally, for $F\neq0$ the memory effects
continue to be present, and the memory kernels $\mu_L(t)$ and
$\mu_R(t)$ for the polymer segments on the left and the right sides of
the membrane are different. In Ref. \cite{anom} we determined
$\mu_L(t)\equiv\mu_{F=0}(t)$ by tethering a polymer of length $N-10$
on a fixed membrane, where we injected $p$ monomers at the tethered
end at time $t=0$, i.e., $v(t)=p\delta(t)$ with $p=10$ (bringing the
final polymer length to $N$), and proxying $\phi(t)$ by the average
distance of the centre-of-mass of the first $5$ monomers $\langle
Z^{(5)}(t)\rangle$ from the membrane. We found
\begin{eqnarray}
\mu_L(t)\sim t^{-\frac{1+\nu}{1+2\nu}}\exp(-t/\tau_{\text{Rouse}})\,,
\label{e4}
\end{eqnarray}
where $\tau_{\text{Rouse}}\sim N^{1+2\nu}$ is the Rouse time, the longest
relaxation time-scale of a polymer of length $N$.
\begin{figure*}
\begin{center}
\begin{minipage}{0.48\linewidth}
\includegraphics[width=\linewidth]{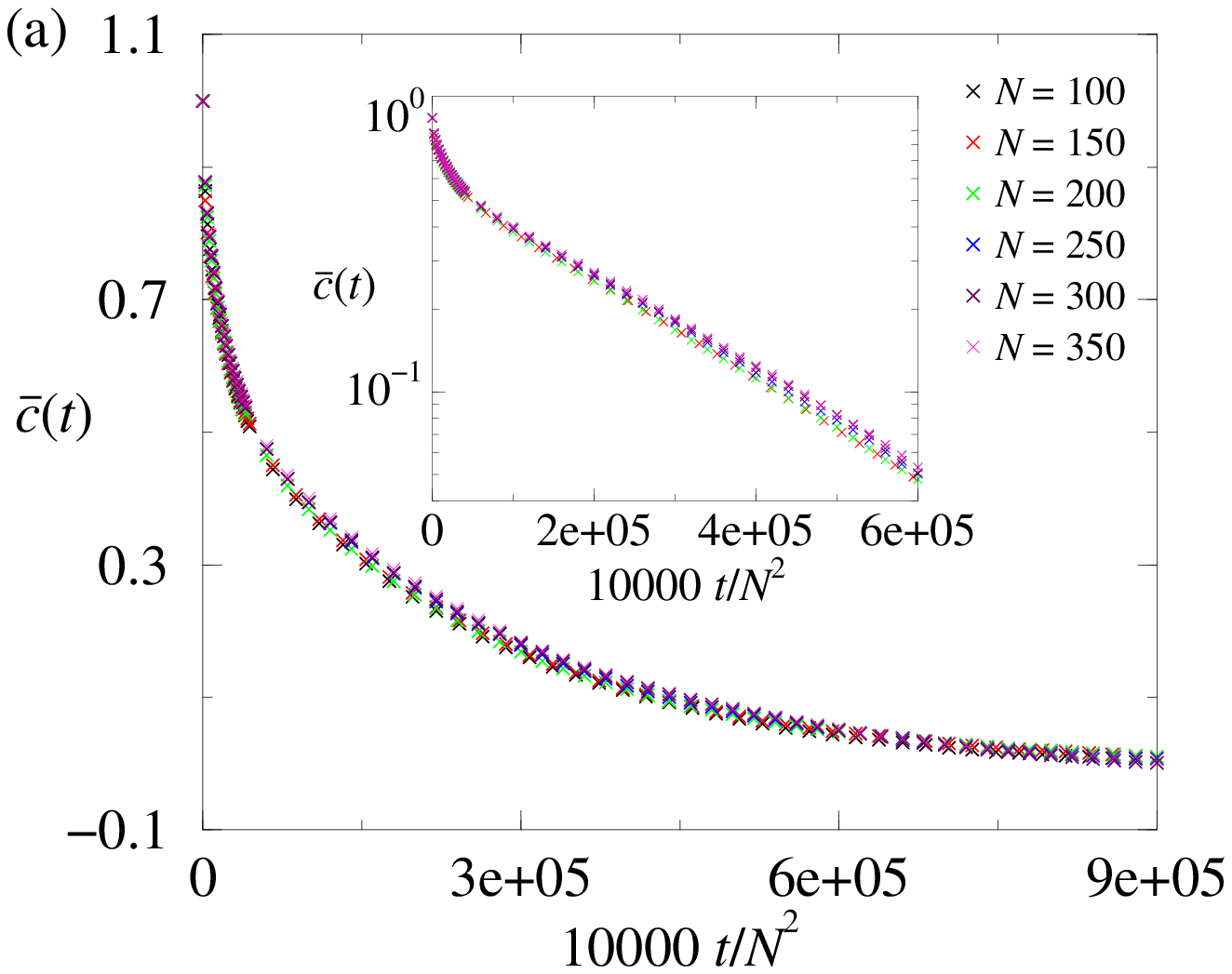}
\end{minipage}
\hspace{5mm}
\begin{minipage}{0.48\linewidth}
\includegraphics[width=0.8\linewidth,angle=270]{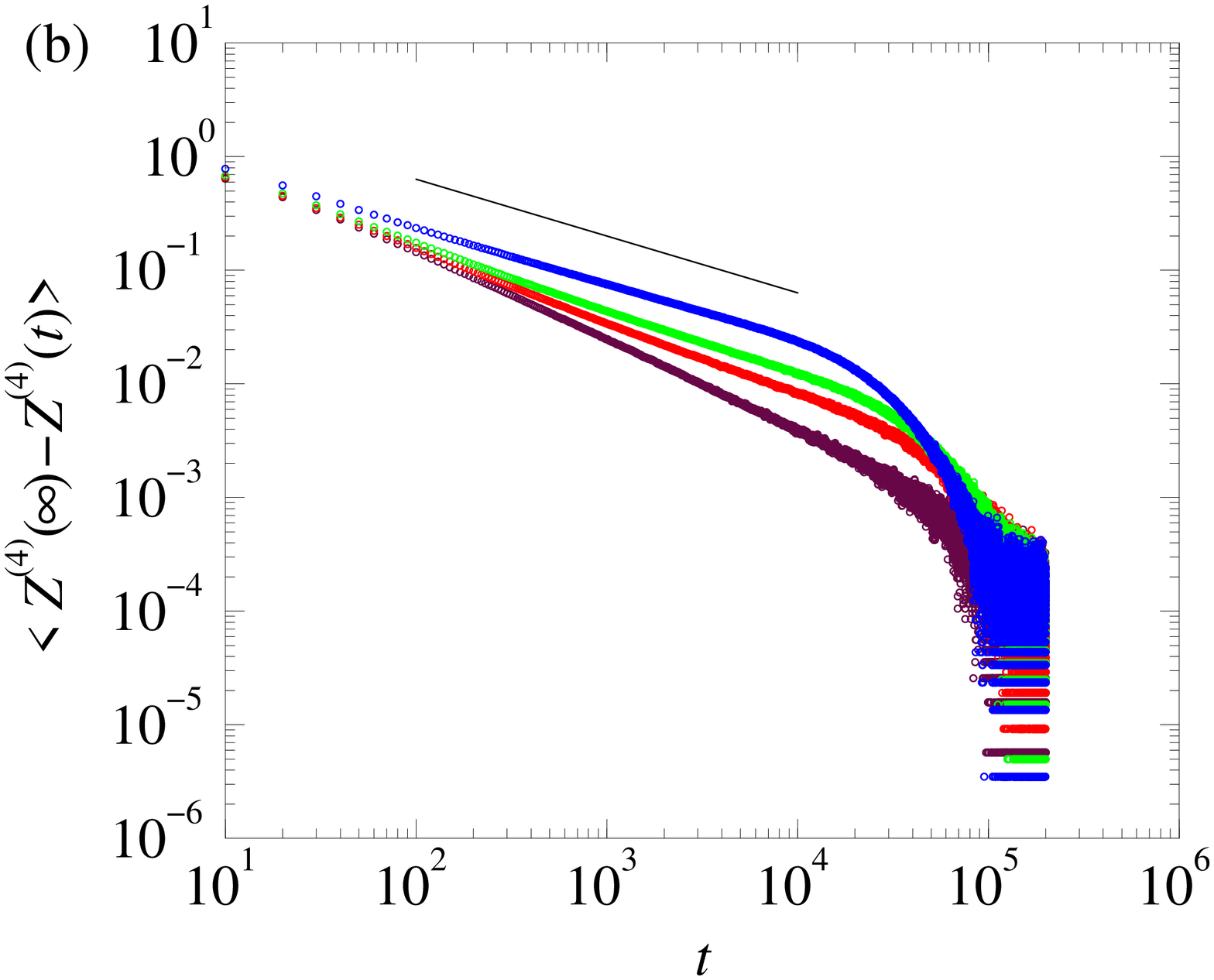}
\end{minipage}
\end{center}
\caption{(a) $\bar{c}(t)\sim\exp(-t/\tau_F)$ for strong forces, with
$\tau_F\sim N^2$; the data shown correspond to $F=1.0$; data
obtained using $256$ polymers for each value of $N$. Inset: the same
data are shown in semi-log plot to show that the decay of
$\bar{c}(t)$ in time is exponential at long times. (b) Behaviour of
$\mu_R(t)$, proxied by $\langle Z^{(4)}(t)\rangle$, for $N=100$ and
four different values of $F$: $F=0.0$ (maroon), $F=0.3$ (red),
$F=0.5$ (green), and $F=1.0$ (blue); the solid black line
corresponds to a slope $t^{-1/2}$; data obtained using $12,800,000$
polymers for each value of $F$. See text for more
details.\label{fig2}}
\end{figure*}

Following the same line as presented in Ref. \cite{anom}, here we
compute $\mu_R(t)$, the memory effect of a polymer of length $N$ with
one end tethered to a membrane, and the other end pulled by a force
$F$. The first step to do this is to obtain the relaxation time for a
polymer of length $N$ under these conditions. For $F=0$ the result for
the relaxation time $\sim\tau_{\text{Rouse}}$ is well-known [and has
been confirmed in an earlier study of ours \cite{anomlong}], but with
$F\neq0$, to the best of our knowledge, the corresponding analytical
result does not exist. We therefore resort to simulations: we denote
the vector distance of the free end of the polymer w.r.t. the tethered
end at time $t$ by $\mathbf{e}(t)$, and define the correlation
coefficient for the end-to-end vector as
\begin{eqnarray}
c(t) =
\frac{\langle\mathbf{e}(t)\cdot\mathbf{e}(0)\rangle-\langle\mathbf{e}(t)\rangle\cdot\langle\mathbf{e}(0)\rangle}{\sqrt{\langle\mathbf{e}^2(t)-\langle\mathbf{e}(t)\rangle^2\rangle\langle\mathbf{e}^2(0)-\langle\mathbf{e}(0)\rangle^2\rangle}}\,.
\label{e4a}
\end{eqnarray}
The angular brackets in Eq. (\ref{e4a}) denote simple ensemble
averaging for $F\neq0$. We first obtain the time correlation
coefficients $c(t)$ for $256$ independent polymers, and $\bar{c}(t)$
is a further arithmatic mean of the corresponding $256$ different time
correlation coefficients. At strong forces, when we scale the units of
time by factors of $N^{2}$ (for self-avoiding polymers!), the
$\bar{c}(t)$ vs. $t$ curves collapse on top of each other. This is
shown in Fig. \ref{fig2}(a) for $F=1.0$ and $N=100,\ldots,350$.

What happens at small to moderate forces to the relaxation time is not
entirely clear to us. We do not expect the relaxation time-scale to
change continuously with $F$. Thus, given the two limits
$\tau_{\text{Rouse}}\sim N^{1+2\nu}$ for $F=0$ and $\tau_F\sim N^2$
for strong forces, we believe that at small to moderate forces the
relaxation time becomes a linear combination of
$\tau_{\text{Rouse}}\sim N^{1+2\nu}$ and $\tau_F\sim N^2$, with the
coefficients of these two times varying with the magnitude of $F$.
\begin{figure*}
\begin{center}
\begin{minipage}{0.48\linewidth}
\includegraphics[width=0.8\linewidth,angle=270]{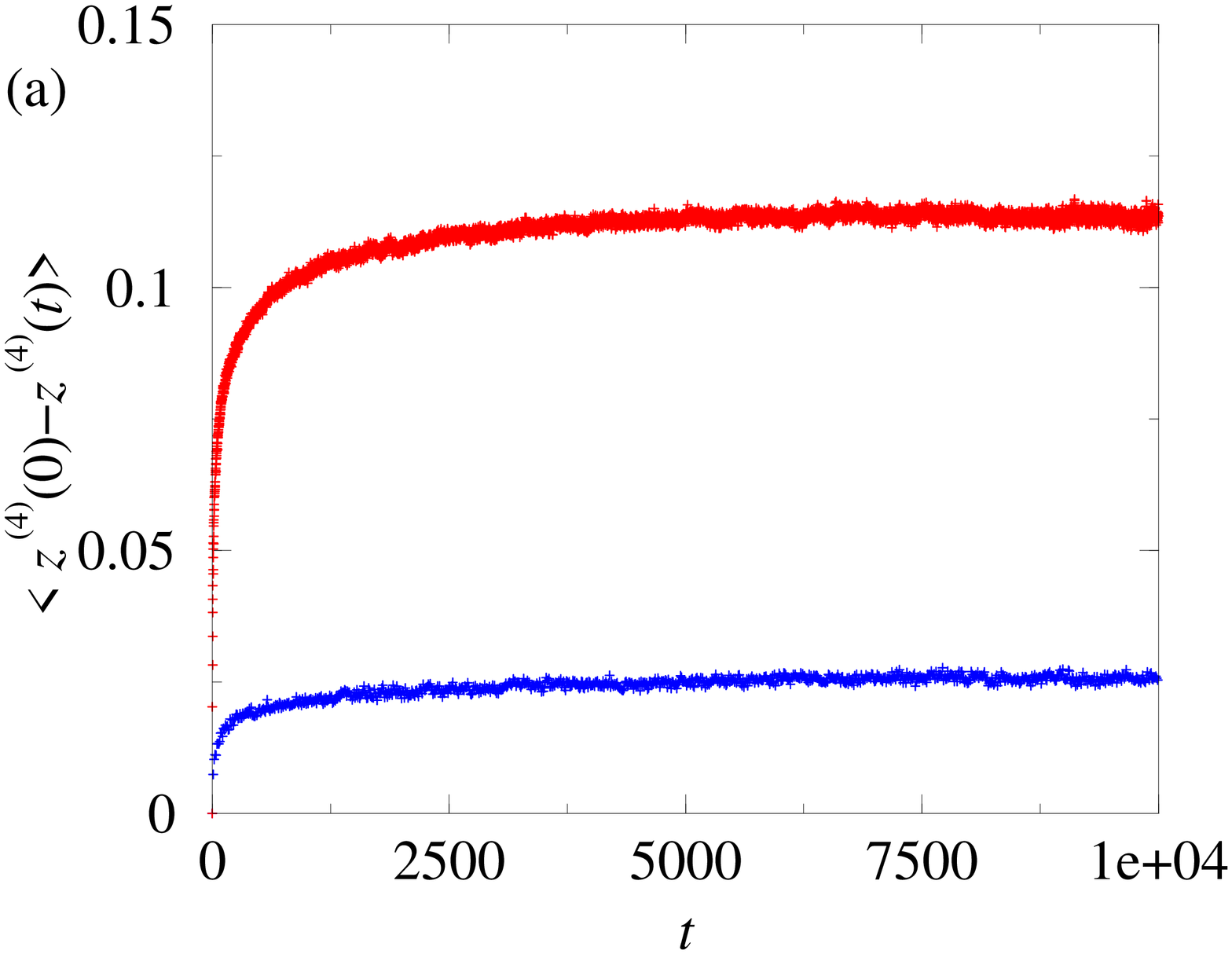}
\end{minipage}
\hspace{0mm}
\begin{minipage}{0.48\linewidth}
\includegraphics[width=0.8\linewidth,angle=270]{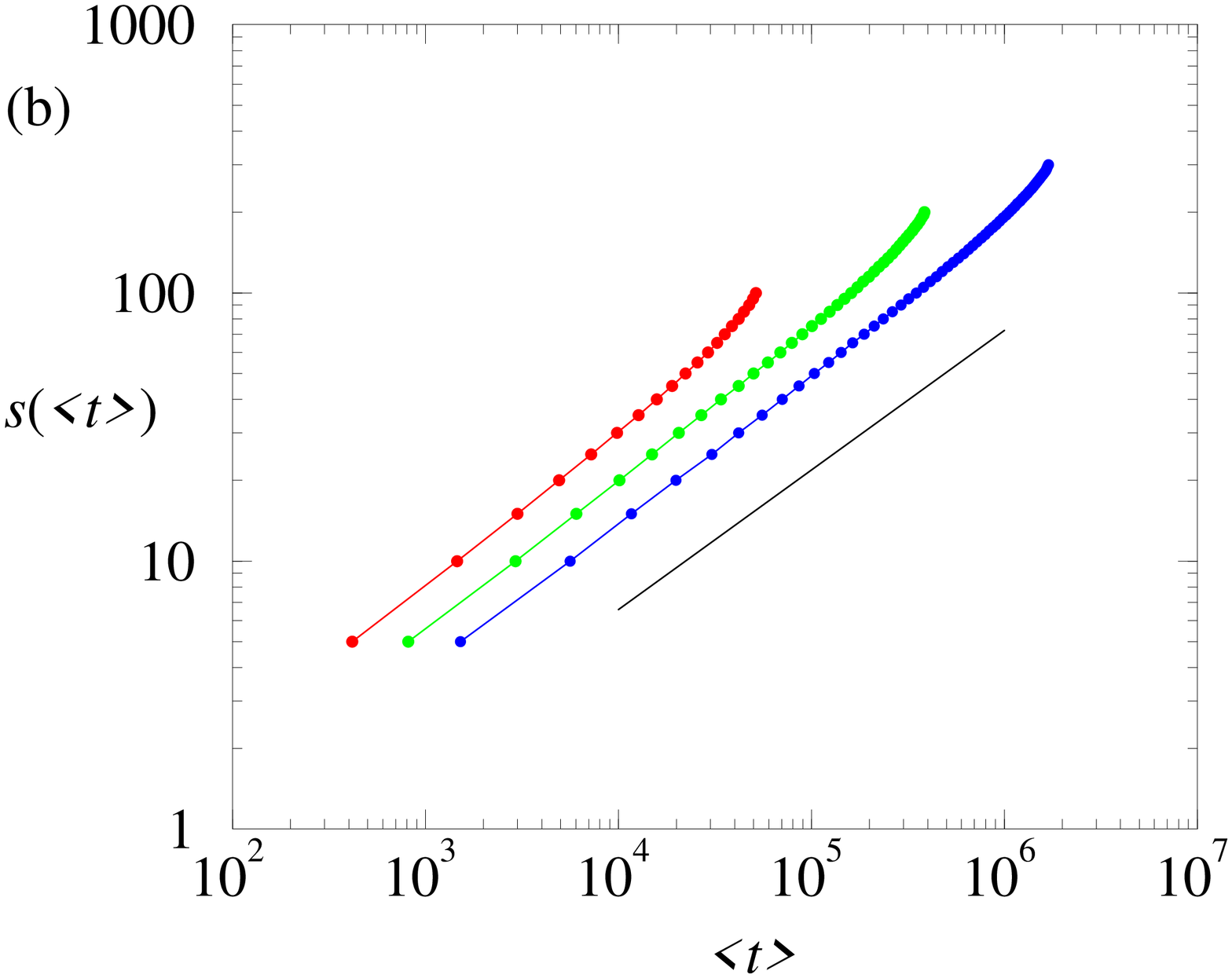}
\end{minipage}
\end{center}
\caption{(a) Behaviour of $\phi(0)-\phi(t)$ for $N=100$ as a function
of $t$, shown by means of the proxy variable $\langle
z^{(4)}(0)-z^{(4)}(t)\rangle$, showing that $\phi(0)-\phi(t)$ reduces
to a constant very quickly: $F=1.0$ (red) and $F=0.5$ (blue). The
angular brackets denote an average over $672,000$ polymer
realizations. (b) Mean time required, for $F=1.0$, to unthread a
distance $s$ for $s=5,10,15,\ldots,N$: $N=100$ (red), $N=200$ (green),
$N=300$ (blue). The time-axis corresponding to $N=200$ is the true
time, for the $N=100$ and $N=300$ cases the time axis is divided and
multiplied by a factor $2$ respectively. This is done in order to show
that the slope of the curves reduces slowly with increasing $N$: we
obtain, for $N=100$ a slope of $0.57$, for $N=200$ a slope of $0.54$,
and $N=300$ a slope of $0.52$ at short times. At long times the slope
increases for all values of $N$: most likely due to the fact that the
monomer at the pore is too close to the end of the polymer. The solid
black line correspond to a slope of $0.52$. The angular brackets denote
an average over $48,000$ polymer realizations. See text for more
details.\label{fig3}}
\end{figure*}

While Fig. \ref{fig2}(a) provides the answer to the relaxation of the
{\it entire\/} polymer for strong forces, the second step to identify
$\alpha$ for $\mu_R(t)\sim t^{-\alpha}\exp(-t/\tau_F)$ for some $\alpha$
for strong forces is to analyze the relaxation of the polymer segments
at the immediate vicinity of the tethered point. The value of $\alpha$
depends on the relaxation properties following the event of injecting,
say, $p$ extra monomers at the tether end, just like extra monomers
add to (or get taken out of) the right segment of the polymer during
translocation. Given the $\exp(-t/\tau_F)$ behaviour of
Fig. \ref{fig2}(a), we anticipate that by time $t$ after the extra
monomers are injected at the tethered point, the extra monomers will
come to a steady state across the inner part of the polymer up to
$n_{t}\sim t^{1/2}$ monomers from the tethered point, but not
significantly further. This internal section of $n_{t}+p$ monomers in
steady state extends only to $r(n_{t})$ from the membrane, because the
larger scale conformation has yet to adjust, and consequently there is
a compressive force $f$ on these $n_{t}$ monomers.

For $F=0$, $r(n_t)\sim n_t^\nu$ is the only length scale for the
equilibrated section of the chain, which leads to $f_{F=0}\sim k_BT\delta
r/r^2$ \cite{anom}, but for $F\neq0$ this does not hold. For small to
moderate forces, i.e., for $FN^\nu/k_BT\lesssim1$, the polymer
conformation is given by a sequence of blobs of size $\xi$, given by
the relation $F\xi=k_BT$ and for strong forces, i.e., for
$FN^\nu/k_BT\gg1$, $\xi\rightarrow a$, where $a$ is the size of a
single monomer \cite{kantor,deGennes}. Thus, for $F\neq0$, the shape
of the polymer resembles that of a cylinder, implying
$f_{F\neq0}\sim k_BT\delta r/(r\xi)$. The independence of $\xi$ on $n_t$
implies that $r(n_t)\sim n_t$, which allows us to write
$f_{F\neq0}\sim k_BT\delta n_t/(n_t\xi)\sim t^{-1/2}$. This force is
transmitted to the membrane, through a combination of decreased
tension at the tether and increased incidence of other membrane
contacts. The fraction borne by reducing tension leads us to what is,
strictly speaking, an inequality: $\alpha\ge1/2$. However, it seems
unlikely that the adjustment at the membrane should be
disproportionately distributed between the two nearly balancing
effects of polymer chain tension and monomeric repulsion, leading to
the expectation that the inequality becomes an equality.

{\it Theoretically\/} however, we cannot rule out the larger values
for $\alpha$, but our numerical results in Fig. \ref{fig2}(b), where
we have used $\langle Z^{(4)}(t)\rangle$ to proxy $\Phi_R(t)$ and
$p=5$, for strong forces favour the smallest theoretical value, namely
$\alpha=1/2$. The power law decay preceding the exponential ones in
Fig. \ref{fig2}(b) change from $t^{-\frac{1+\nu}{1+2\nu}}$ at $F=0$ to
$t^{-1/2}$ for strong forces ($FN^\nu/k_BT\gg1$). Following the
discussion about relaxation times for small to moderate forces three
paragraphs above, we believe that between $F=0$ and $FN^\nu/k_BT\gg1$
the decay for $\Phi_R(t)$ at short times is a combination of these two
power laws. Additionally, closer inspection of the curves for $F=0.3$
and $F=0.5$ in Fig. \ref{fig2}(b) reveals that the slope is steeper in
the beginning: perhaps it is an indication that relaxation within a
blob (corresponding to $t^{-\frac{1+\nu}{1+2\nu}}$) precedes
inter-blob rearrangements (corresponding to $t^{-1/2}$). Nevertheless,
at strong forces, the behaviour
\begin{eqnarray}
\mu_R(t)\sim
t^{-\frac{1}{2}}\exp(-t/\tau_F),\quad\mbox{with}\quad\tau_F\sim N^2,
\label{e4b}
\end{eqnarray}
stands as a witness of the fact that the Flory-like structure of the
polymer is entirely destroyed.

\section{The relation between $\phi(t)$ and 
  $v(t)$, and the scaling behaviour of $\tau_N$\label{sec3}}

\subsection{Relation between the imbalance of chain tension $\phi(t)$
  and the translocation velocity $v(t)$ \label{sec3.2}}

In this subsection we consider the strong force case as it is
simpler. The moderate to weak force case is discussed in
Sec. \ref{sec3.3}.

So far, we have $\mu_L(t)\sim
t^{-\frac{1+\nu}{1+2\nu}}\exp(-t/\tau_{\text{Rouse}})$ and
$\mu_R(t)\sim t^{-1/2}\exp(-t/\tau_F)$ for strong forces. Since the
memory effects in the dynamics of the translocating polymer stem from
the power laws, in the absence of symmetry between the left and the
right segment of the polymer, we only need to keep track of the power
law of $\mu_R(t)$, as it has a {\it lower\/} exponent than
$\mu_L(t)$. In other words, in the relation
\begin{eqnarray}
\phi(t)=\phi_{t=0}-\int_{0}^{t}dt'\,|\mu(t-t')|\,v(t')\,,
\label{e5}
\end{eqnarray} 
we have to use the fact that the power law decay of $\mu(t)$ behaves
$\sim t^{-1/2}$. Note the absolute value around $\mu(t)$, as the sign
of $\mu(t)$ is negative.

Equation (\ref{e5}) can be inverted via Laplace transformation,
yielding
\begin{eqnarray}
v(k)=\frac{\phi_{t=0}}{k|\mu(k)|}\,-\,\frac{\phi(k)}{|\mu(k)|}\,,
\label{e6}
\end{eqnarray} 
where $k$ is the Laplace variable representing inverse
time. Thereafter, using $\mu(t)\sim t^{-1/2}$, i.e., $\mu(k)\sim
k^{-1/2}$, and Laplace-inverting Eq. (\ref{e6}), we get
\begin{eqnarray}
v(t)=\int_{0}^{t}dt'\,(t-t')^{-3/2}\,[\phi_{t=0}-\phi(t)]\,.
\label{e7}
\end{eqnarray} 

\subsection{Scaling behaviour of $\tau_N$ with $N$ \label{sec3.3}}

In Eq. (\ref{e7}), if $\phi(t)$ goes to a constant $\neq\phi_{t=0}$,
then
\begin{eqnarray}
v(t)\sim t^{-1/2}\quad\mbox{i.e.,}\quad
s(t)=\int_{0}^{t}dt'\,v(t')\sim t^{1/2}\,,
\label{e8}
\end{eqnarray} 
where $s(t)$ is the distance unthreaded in time $t$ \cite{note1}.

In Fig. \ref{fig3}(a) we show the behaviour of $[\phi_{t=0}-\phi(t)]$
by means of the proxy variable $\langle z^{(4)}(0)-z^{(4)}(t)\rangle$
for strong forces [$F=1.0$ (red) and $F=0.5$ (blue)], where
$z^{(4)}(t)$ is the difference between the $Z^{(4)}(t)$ values between
the right and left segment of the polymer, i.e.,
$z^{(4)}(t)=Z^{(4)}_R(t)-Z^{(4)}_L(t)$. Indeed $[\phi_{t=0}-\phi(t)]$
goes to a constant fairly quickly. Following Eq. (\ref{e8}), this
yields us the scaling $s(t)\sim t^{1/2}$ for strong forces. The data
in support of the scaling $s(t)\sim t^{1/2}$ are shown in
Fig. \ref{fig3}(b), for $F=1.0$.

\begin{figure}
\begin{center}
\includegraphics[width=\linewidth]{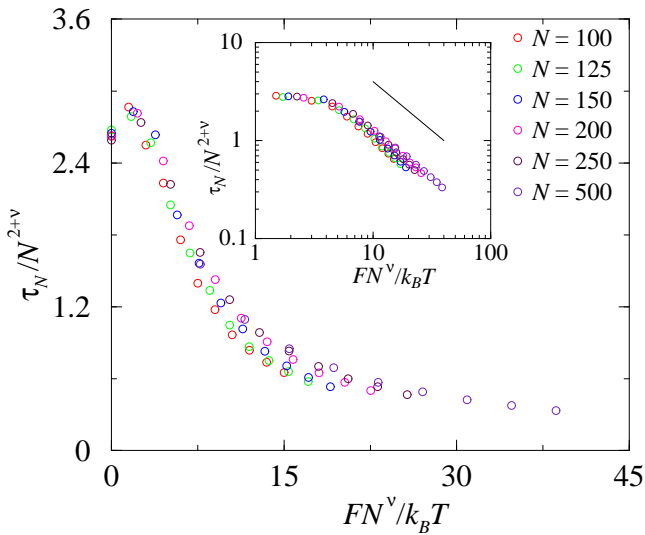}
\end{center}
\caption{Collapse of all data in terms of $FN^\nu/k_BT$ and
$\tau_N/N^{2+\nu}$ for $F=0.0,0.1,0.2,\ldots,1.0$. Inset: the same
data in log-log plot, the black line corresponds to a slope of
$-1$. The $\tau_N$ values correspond to the median for $1,024$
polymer unthreading events. \label{fig4}}
\end{figure}

The scaling for the mean unthreading time $\tau_N$ is obtained from
the equation $s(\tau_N)=N$. For strong forces, it is derivable as
$\tau_N\sim N^2$ --- as shown in Fig. \ref{fig4}(a) or in earlier
works \cite{kantor,luo} --- from Eq. (\ref{e8}) if we assume that
$[\phi_{t=0}-\phi(t)]$ is a constant {\it independent\/} of $N$. It
seems reasonable (and likely!) that a local property like
$[\phi_{t=0}-\phi(t)]$ should be unaffected by the polymer length,
which is a large-scale property; nevertheless, we have no way to argue
this theoretically. In the context of using Eq. (\ref{e8}) to obtain
$\tau_N\sim N^2$, it is however useful to note that in the scaling
sense $\tau_N$ is smaller than (or equal to) the time scales in the
exponential decay of $\mu_R(t)$ and $\mu_L(t)$, otherwise the
power-law behaviour of $\mu(t)$ we used in Eq. (\ref{e5}\ref{e8})
would not have been applicable for all times $t<\tau_N$.

The collapse of all the data for the unthreading times for several
different values of $N$ and $F$ in terms of the variables
$FN^\nu/k_BT$ and $\tau_N/N^{2+\nu}$, as shown in Fig. \ref{fig4},
indicates that the unthreading time $\tau_N$ can be written in a
scaling form as
\begin{eqnarray}
\tau_N\sim N^{2+\nu}\,\,g\left(\frac{FN^\nu}{k_BT}\right)\,,
\label{e9}
\end{eqnarray} 
where $g(x)$ is a scaling function of its argument $x$. Figure
\ref{fig4} shows that $g(x)\sim 1/x$ for $x\gg1$. Note that
$g(0)$ is a constant, in agreement with our earlier result that the
unthreading time scales as $N^{2+\nu}$ in for unbiased translocation
\cite{anom}, in which case the polymer leaves the pore purely due to
thermal fluctuations.

The fact that $g(0)$ is a constant indicates that $g(x)$ has to
deviate from the $1/x$ behaviour as $x$ approaches zero. From the
inset of Fig. \ref{fig4} we see that $g(x)$ starts to deviate from
the $1/x$ behaviour at about $x=4$, at which point the force is
moderate in strength. In Fig. \ref{fig4}, note also that
$\tau_N/N^{2+\nu}$ has a higher prefactor close to $x=0$ than at $x=0$.

{\it A priori\/}, the same analysis (\ref{e5}-\ref{e8}) holds for
small to moderate forces as well. Nevertheless, whether Eq. (\ref{e7})
is actually useful in such circumstances is a different
matter. Indeed, a deeper investigation reveals that $\phi_{t=0}$ for
small to moderate forces can be extremely small. To give a feeling for
how small $\phi_{t=0}$ can be, we obtained $\langle
Z^{(4)}(t=0)\rangle$ values for $N=100$ for $F=0.0,0.1,\ldots,1.0$
(not all are plotted in Fig. \ref{fig1}). The $\langle
Z^{(4)}(t=0)\rangle$ values for $F=0.1,\ldots,0.3$ (approximate
$x$-values $1.5,3$ and $4.5$ in Fig. \ref{fig4}), corresponding to the
right segment of the polymer, turned out to be $1.35, 1.36$ and $1.38$
respectively, while $\langle Z^{(4)}(t=0)\rangle$ corresponding to
$F=0.0$ (i.e., for the left segment of the polymer) turned out to be
$1.34$.

Since $FN^\nu/k_BT$ is a dimensionless parameter that describes the
effect of the force on the polymer's dynamics compared to the effect
of thermal fluctuations, it seems logical that if $FN^\nu/k_BT$ is slowly
reduced, thermal fluctuations start to dominate over the effect of the
force, and translocation by the pulling force $F$ starts to resemble
unbiased translocation, i.e., translocation in the absence of any
external forces \cite{wolt,anom,anomlong}. Such a picture is
manifested by both sides of Eq. (\ref{e7}) effectively becoming zero as
suggested in the above paragraph; the equation remains valid, but
ceases to be useful in practice.

\section{Discussion\label{sec4}}

In this paper, we have considered polymer translocation pulled through
a narrow pore by a force $F$. We have provided a theoretical
description of the polymer's dynamics under these conditions, as well
as of the scaling behaviour of the unthreading time $\tau_N$ for the
polymer of total length $2N$, the time the polymer takes to leave the
pore. Our theory is supported by high precision computer simulation
data, generated for a three-dimensional self-avoiding lattice polymer
model.

At strong forces, i.e., for $FN^{\nu}\gg1$, we have reported $\tau_N\sim
N^2/F$: we have shown that the translocation velocity $v(t)$ is {\it not\/}
constant in time; in fact, the velocity of translocation $v(t)$ is
shown to behave as $t^{-1/2}$, while for small to moderate forces the
behaviour of $v(t)$ is more complicated. At small to moderate forces,
satisfying the condition $FN^{\nu}\lesssim1$, where $\nu\approx0.588$
is the Flory exponent for the polymer, we have found that $\tau_N$
is independent of $F$, and in agreement with our earlier result for
unbiased polymer translocation \cite{anom} scales with polymer length as
$\tau_N\sim N^{2+\nu}$. 

We have shown that the scaling of $v(t)$ as well as the $N$-dependent
part of $\tau_N$ stem from the dynamics of the polymer segments at the
immediate vicinity of the pore --- in particular, the memory effects
in the polymer chain tension imbalance across the pore. The
theoretical analysis presented here is based on that of
Ref. \cite{anom}, and therefore provides a direct confirmation of the
robustness of the theoretical method presented in
Ref. \cite{anom}. Additionally, we note that the physical picture
provided in Refs. \cite{kantor,luo}, wherein  the scaling arguments
for the unthreading time involved a constant velocity of translocation
(albeit an average one, in light of this work) is incomplete.

It should nevertheless be mentioned that the dependence of the
relevant quantities, such as $v(t)$ or $\tau_N$ on $F$ is beyond the
scope of the theoretical description provided here. The main reason
behind this is that no analytical expression has been reported
(neither do we have one ourselves) for the quantities, such as the
polymer chain tension, memory kernel etc. for force $F$. Indeed, the
behaviour of the quantities of interest on $F$ is complicated, as
already manifested by Fig. \ref{fig1}, and in the absence of a
theoretical description involving $F$, numerical investigation has
remained the only way. Nevertheless, we note that the $y$-axis of
Fig. \ref{fig4} as $\tau_N/N^{2+\nu}$ [originating from our previous
work \cite{anom}], the $x$-axis of Fig. \ref{fig4} as $FN^\nu/k_BT$ as
a measure of the strength of the force in relation to thermal
fluctuations, and the scaling $\tau_N\sim N^2$ at strong forces
automatically imply that $\tau_N$ has to behave $\sim1/F$ at strong
forces.

\section*{}

{\bf Acknowledgements:} We gratefully acknowledge our discussions with
Prof. Robin C. Ball. Virtually unlimited computer time from the Dutch
National supercomputer cluster SARA is also acknowledged.


\begin{thebibliography}{10}
\bibitem{evi} There is however numerical evidence in Ref. \cite{luo}
  that the velocity of translocation is not constant in time (their
  Fig. 3).
\bibitem{note1} Note that with $\phi(t)$ a constant, strictly
  speaking, the integral (\ref{e7}) does not converge. The divergence
  stems from the assumption that $\mu(t)\sim t^{-1/2}$ holds all the
  way to $t\rightarrow 0$. This is clearly not true as can be seen
  from Fig. \ref{fig2}(a), which provides the required cutoff for the
  convergence of the integral (\ref{e7}).
\end{thebibliography}
\end{document}